\documentclass[a4paper]{iopart}

\usepackage{latexsym} 
\usepackage{amssymb}
\usepackage{color}
\usepackage{ulem}
\usepackage{times}
\usepackage{colordvi}

\usepackage{ifpdf}
\usepackage{graphicx}
\newcommand{\imgname}[1]{#1.eps}

\ifpdf
\fi
\ifpdf
  \renewcommand{\imgname}[1]{#1.pdf}
  
\fi
\begin{document}

\title[Local dark matter searches with LISA]{
       Local dark matter searches with LISA}

\author[M. Cerdonio et al.]{
        Massimo Cerdonio$^{1}$, 
        Roberto De Pietri$^{2}$, 
        Philippe Jetzer$^{3}$ and 
        Mauro Sereno$^{3}$}

\address{$^1$ Dipartimento di Fisica, Universit\`a di Padova, Italy and 
              INFN sezione di Padova, Italy}

\address{$^2$ Dipartimento di Fisica, Universit\`a di Parma, Italy  and 
              INFN, gruppo di Parma, Italy}

\address{$^3$ Institute of Theoretical Physics, University of Z\"urich, Switzerland}

\date{\today}
\begin{abstract}
The {\it drag-free} satellites of LISA will maintain the test masses in
geodesic motion over many years with residual accelerations at
unprecedented small levels and {\it time delay interferometry} (TDI) will
keep track of their differential positions at level of
picometers. This may allow investigations of fine details of the
gravitational field in the Solar System previously inaccessible.

	In this spirit, we present the concept of a method to measure
directly the gravitational effect of the density of diffuse Local Dark
Matter (LDM) with a constellation of a few {\it drag-free} satellites, by
exploiting how peculiarly it would affect their relative motion.

	Using as test bed an idealized LISA with {\it rigid} arms, we find
that the separation in time between the test masses is uniquely
perturbed by the LDM, so that they acquire a differential breathing
mode. Such a LDM signal is
related to the LDM density within the orbits and has characteristic
spectral components, with amplitudes increasing in time, at various
frequencies of the dynamics of the constellation. This is the
relevant result, in that the LDM signal is brought to non-zero
frequencies.

\end{abstract}

%\maketitle

%-----------------------------------------------------------------
\section{Introduction}
\label{sec:intro}
 %-----------------------------------------------------------------

According to the current cosmology paradigm dark energy and dark matter are
necessary to understand the currently observed cosmological expansion and as
well dark matter appears necessary to understand  galactic scale
phenomenology \cite{weinberg}. In particular the flat rotation curves of galaxies
indicate that, at galactic scales, the dark matter can be some five orders
of magnitude greater than the average cosmic value.

High precision Solar system tests have been providing model independent constraints 
on the local dark matter. The Solar system is the larger one with very well known mass 
distribution and can offer tight confirmations of Newtonian gravity and general relativity. 
What is usually investigated is the gravitational action of dark matter. Experimental bounds 
on non luminous matter in Solar orbit were derived either by considering the third Kepler's 
law \cite{and+al89,and+al95} or by studying its effect upon perihelion precession \cite{gr+so96}. 
The influence of a tidal field due to Galactic dark matter on the motion of the planets and satellites 
in the Solar system was further investigated by \cite{bra+al92} and \cite{kl+so93}. The orbital motion 
of Solar system planets has been determined with higher and higher accuracy \cite{pit05b} and recent 
planetary astrometric data allowed to put interesting limits. The most recent analyses 
give an upper limit on the local dark matter density of 
$\rho_\mathrm{DM} < 3 {\times} 10^{-16}~\mathrm{kg/m}^3$ 
at the 2-$\sigma$ confidence level \cite{1,2,3,4}. 
Such a limit falls short to estimates from Galactic dynamics \cite{5} by 5-6 orders of magnitude. 
Future radio ranging observations of outer planets with an accuracy of few 
tenths of a meter could either give positive evidence of dark matter or disprove modifications of gravity.

We present here, in
connection with the dynamics of the drag-free LISA constellation, a
new method, which has the advantage to bring the secular effects to
non-zero frequencies and thus to considerably ease possible detection. 

If applied to a direct gravitational
measurement of the total LDM density in the Solar System, 
the method would be quite relevant, as  it would
allow to make a comparison with the presence/absence of
the proposed components of local dark matter, the search for which is
actively under way in a variety of direct laboratory experiments 
and indirect space based astrophysical observations \cite{6}. 
In the present paper such a
comparison at present will be made taking into consideration the
diffuse GDM density, taken as uniformly distributed, but there is
the possibility that the local value of the LDM at the Solar System may
be somewhat \cite{7} or even considerably higher \cite{8},
although recent numerical simulations indicate that the value of the LDM density in the Solar 
System should be quite close to that of the GDM diffuse density \cite{vogel}.

Before going to the details we notice that our method
does not depend on the nature of the dark matter, whether it is baryonic
or non-baryonic and behaves as cold, warm or hot dark matter as long
as the dark matter has the usual gravitational interaction with 
ordinary matter. 
We assume that the dark matter is a homogeneous diffuse background of elementary particles
(as the candidates which are searched for in the underground experiments
such as WIMPs (the leading candidate for which is a neutralino) and axions) or at most small
clumps with mass much smaller than the total dark matter mass contained
in a sphere with radius of the order of the LISA orbit of about 1 Astronomical
Unit, where we assume an average dark matter density equal to the LDM density.
Clearly, we are mainly thinking in our proposal of dark matter in form of
elementary particles as WIMPs.So our study is somewhat complementary to the 
analysis given in \cite{danzmann} for the impulsive disturbances LISA would suffer from the 
close encounter with dust grains and small bodies belonging to the upper mass range 
of interplanetary dust, with masses of $10^{-3}$ kg to $10^{15}$ kg, which would 
apply quite the same to hypothetical dark matter clumps of that masses. Similarly
if the dark matter is in the form of small-mass primordial black holes in the range
 of $10^{11}$ kg to $10^{17}$ kg then it has been shown that a nearby encounter of such 
an object with one of the LISA spacecrafts would lead to a detectable 
pulse-like signal \cite{seto,adams}. A difficulty is then to distinguish 
this class of events with those involving perturbations due to
the close encounter of near-Earth asteroids and large dust grains.

%-----------------------------------------------------------------
\section{The method}
\label{sec:method}
%-----------------------------------------------------------------

We use as test bed a simplified version of the planned LISA mission,
which well serves the purpose of studying the basic concept. To lowest
order in the eccentricities, it has been shown that the LISA $5\times 10^6$ km
triangular flight formation, with given choices for the heliocentric
orbits, is stable and rotates rigidly around the centre of symmetry,
which in turn executes a yearly revolution around the Sun, trailing
behind the Earth on the same orbit \cite{9}. We take this approximated motion
of the LISA constellation as the one to be perturbed by a local
diffuse matter density. In this approximation both the
corrections due to higher terms in the eccentricity and all
perturbations due to the Earth, the Moon, and other planets are
disregarded, and as well as all the corresponding coupling effects, while
in reality the constellation is expected to breathe yearly by
thousands of km because of them. Actually such a large breathing is already
present for the exact Keplerian orbits of ref. \cite{9}
in the field of the Sun alone.

To calculate the motion of a test mass in presence of an homogeneous 
background in addition to a central mass $M$ we can use \cite{10} the Newtonian 
expression 
\begin{equation}
\label{poi2}
\phi = -\frac{G M}{r} - \frac{k}{2}r^2~,
\end{equation}
where $G$ is the gravitational constant and $r$ the distance from the central mass.
The second term in the
potential of Eq.~(\ref{poi2}) 
gives to a test mass in orbit around the Sun a
perturbing central acceleration of the kind $\sim k\mathbf{r}$, with 
$\mathbf{r}$ the vector
distance from the Sun, $r=1 \mathrm{AU} = 1.5\times 10^{11}$ m for LISA.
We remind that the for an homogeneous background the choice
of the centre of the coordinate system does not affect the result and
thus we can take the Sun as the origin of it \cite{xx}.

Within the current cosmological paradigm the constant $k$ would be the sum
of three terms of different amplitude and sign, respectively one proportional
to the density $\rho_{LDM}$ of LDM, the second one proportional to the density of cosmological dark matter
and the third one related to the dark energy.
As remarked above, according to current views the LDM term is at least five orders
of magnitude higher than the other two terms, which therefore we disregard.
In any case this shows that in principle the signals we shall discuss here would
carry also information about cosmological dark matter/energy, but as 
it stands 
now the cosmological contributions appear to be masked by many orders of 
magnitude by the local dark matter effects. 
Therefore for the constant $k$ we take 
$k = - 4 \pi G \rho_\mathrm{LDM}/3$.
 
The $k\mathbf{r}$ perturbation can be addressed with
standard Lagrange's planetary equations for the evolution of orbital
elements \cite{11}. In this framework the orbits keep on
average the Keplerian behavior, but in detail there is a yearly
modulation of orbital elements and it comes about a secular precession
of perihelion. Lagrange's planetary 
equations account for the time evolution of the orbital elements of the 
Keplerian orbit. Let us denote the semi-major axis length, 
the eccentricity and the periastron argument with $a$, $e$ and $\omega$, 
respectively. Since the perturbation $k \mathbf{r}$ is radial,  
the inclination and the longitude of ascending node do not change, whereas the other orbital elements evolve as
\begin{eqnarray}
\frac{d a}{d t} &  = & \frac{2}{n}\frac{e}{\sqrt{1-e^2}}~ k r \sin \varphi , \\
\frac{d e}{d t} &  = & \frac{1}{n a} \sqrt{1-e^2}~ k r \sin \varphi , \\
\frac{d \omega}{d t} &  = & - \frac{1}{n a}\frac{\sqrt{1-e^2}}{e}~ k r \cos \varphi  ,
\end{eqnarray}
where $\varphi$ is the mean anomaly and $n \equiv \sqrt{G M/a^3}$ is the unperturbed Keplerian mean motion. 
Solving the Lagrange's equations, we get the evolution up to terms linear in the parameter $k$. 
The above equations can be solved keeping the orbital elements on the right-hand side unperturbed. Since the orbits of LISA satellites are going to be nearly circular we can also expand in the eccentricity.

%%%----------------------------------------------------------------
\begin{figure}[t]
\begin{flushright}
\includegraphics[width=0.85\textwidth]{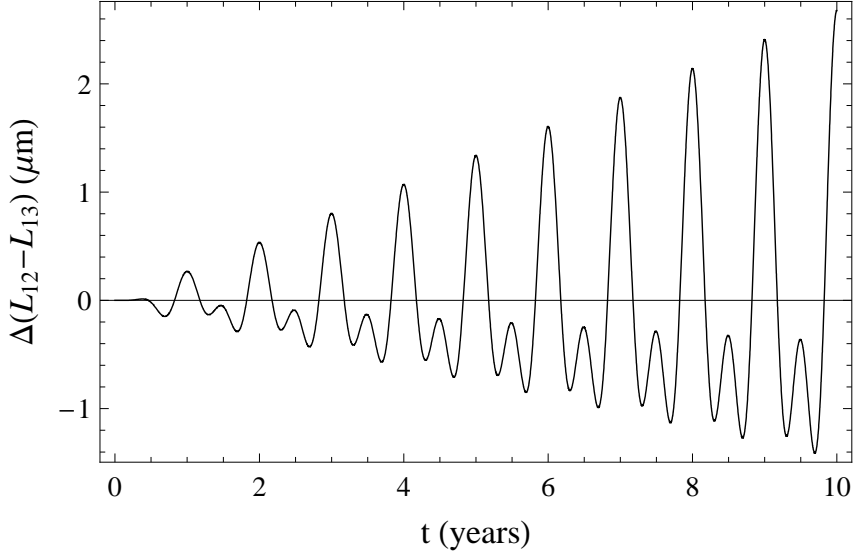}
\end{flushright}
\vglue-0.1cm
\caption{Differential motion due local dark matter between one pair of
LISA arms after release in geodesic motion, as from the time
dependence of arm lengths ($L_{12}$ is the distance between test
masses in spacecrafts \# 1 and \# 2 and similarly $L_{13}$); the
differential motions between the other pairs are similar. The value assumed in the 
calculations for the diffuse LDM density is 0.3 GeV cm$^{-3}$.}
\label{fig:spacecraft}
\end{figure}

As a worked example we study the effect, as LDM,  
of the accepted average GDM density of $\rho = 0.3$ GeV/cm$^3$ \cite{5},
assumed to be diffuse and uniformly distributed as such within the Solar
System. This gives $k = -1.5\times 10^{-31}$ s$^{-2}$ on the idealized
rigid LISA orbits of ref. \cite{9}. The $k\mathbf{r}$ perturbation
operates on each of the test masses of LISA, which execute Keplerian
orbits to realize the rotating triangle constellation. What is
relevant here is the temporal behavior of the arm lengths
($L_{ij}$) and of the differences between arm lengths, $\Delta L_{ij}-\Delta L_{is}$. 

The lowest approximation in eccentricity gives arm lengths $L_0$ that are time
independent. If we now insert the perturbation,
when we look for the
relative motions between the test masses, we find that, after
release in geodetic motion, the arms acquire under the perturbation a
breathing mode, with characteristic spectral components both at the
fundamental $1 y^{-1}$ frequency of revolution around the Sun, and 
higher frequency contributions including the 
$3 y^{-1}$ frequency with which the LISA constellation returns to
itself in its rotation around its barycentre.
Most importantly, in the prospect of detection with TDI methods, we
find
\begin{eqnarray}
\Delta (L_{12} - L_{13}) &=&
-\frac{3 \sqrt{3} L_0}{16 n^2} k
 \bigg[n t \bigg(4 \cos (n t)+3 \cos (2 n t)\bigg)
\label{eq9}\\
 &&+5 \sin (nt)-\frac{11}{2} \sin (2 n t)-\frac{1}{3} \sin (3 n t)
 \bigg]
  \nonumber
\end{eqnarray}
a time dependence of the difference between the changes in length
of pairs of arms (see Eq.~(\ref{eq9}) and Fig.~\ref{fig:spacecraft}). 
Even more interestingly, such an amplitude modulation increases with
time, giving thus to the signal a characteristic time-frequency
signature. We also checked the accuracy of these results by a direct 
numerical integration of the equation of motion finding 
a perfect agreement between numerical and analytical results.
   
A differential motion of the order of micrometers, as in
Fig.~\ref{fig:spacecraft}, may appear comfortably large in respect to
TDI capabilities, but one should take notice of the low frequencies at
which it occurs. A realistic feeling of numbers comes about as
follows. The performance of LISA, in regard to the lowest residual
acceleration in respect to geodesic motion, could be at best $3\times
10^{-15}$ m s$^{-2}$ Hz$^{-1/2}$ at $10^{-4}$ Hz and possibly $10^{-13}$
m s$^{-2}$ Hz$^{-1/2}$ at $10^{-6}$ Hz \cite{12a}, which, over a $3 y$
integration time, translates into $3\times 10^{-19}$  m s$^{-2}$ and
$10^{-17}$ m s$^{-2}$ rms noise level respectively. These numbers should
be compared with the value reached by the differential acceleration
between the spacecrafts $k r \simeq 5 \times 10^{-20}$ m s$^{-2}$ given by the GDM
in $3 y$ of its secular increase. Of course if the LDM density would
be much higher than the GDM, as proposed in ref. \cite{8}, an actual
detection by LISA might appear in principle possible, but still,
within the oversimplified model we analyzed, it would appear at
characteristic frequencies not higher than $10^{-7}$ $\mathrm{Hz}$, frequencies
at which we cannot expect that much from LISA in any case.

In conclusion, we see that the effect of LDM shows in the presence 
of distinctive signals on the breathing dynamics that sheds light on the
possibility, in principle, of detecting, or putting upper limits, on 
the gravitational action of LDM. However, at this stage, it is not clear
if such a method would be competitive with respect to planetary upper 
bounds \cite{1,2,3}. As we discuss in the next section,
an LDM signal power may appear at distinctive frequencies
where the un-perturbed contributions are small or zero. 
In such a framework, a few problems arise which
we discuss in the next section together with their possible solutions.

%-----------------------------------------------------------------
\section{Discussion}
\label{sec:Discussion}
%-----------------------------------------------------------------

First, even in absence of other perturbations, when the rigid motion
around the Sun of LISA used above is taken to higher approximation in
the eccentricity \cite{9}, the arms acquire breathing modes of large
amplitude. We have made analytical calculations up to 4th order in
eccentricity respectively both without and with the perturbation due 
to LDM for the differential breathing of the arms
$\Delta L_{ij}-\Delta L_{is}$. We find that:

\begin{itemize}
\item {\it without} the LDM perturbation {\it i)} there is signal power at
      the fundamental $1\mathrm{y}^{-1}$ and overtones only up to $7 \mathrm{y}^{-1}$,
      {\it except} $3 \mathrm{y}^{-1}$ and $6 \mathrm{y}^{-1}$; 
      {\it ii)} the amplitude of the modes {\it does not} increase in time
\item {\it with} the LDM perturbation {\it i)} there is signal power at the
       fundamental $1y^{-1}$ and at all overtones up to $11\mathrm{y}^{-1}$, {\it including}
       $3\mathrm{y}^{-1}$ and $6\mathrm{y}^{-1}$; {\it ii)} the amplitude of all the modes 
       {\it does} increase in time.
\end{itemize}

\noindent
So it appears that the LDM signal has a specific signature
in the spectrum. 

Second we must take into account the perturbation of other
matter present in the Solar System, that is planets, asteroids,
interplanetary dust, Solar wind. As for the gravitational effect of a
local density of particles, atoms and dust, both in its steady value
and in its variability due to Solar activity, it should be possible to
take them in account accurately.

The distribution of the dust density is supposed to be flat
centred on the Sun, with cylindrical symmetry and of thickness of the order 
of 1 AU, extending to some 3 AU \cite{12}. The value of its total density at Earth 
orbit, but far from the Earth so to ignore its gravitational pull, is indicated as 
$\rho_\mathrm{dust}$  $(1 AU) = 9.6 \times 10^{-20}$ kg/m$^3$ \cite{13}. We have to estimate the 
effect of dust and compare to that of LDM, as they may come out to be 
similar and hard to disentangle. The LISA constellation in its orbit stays well 
within an approximately uniform distribution of dust at the density value 
quoted above and we assume to be the same for the LDM distribution, so that 
we may consider LISA embedded in a spherically symmetric density of dust and LDM. 
This will be an extremely crude estimate, giving a conservative upper bound as the dust 
distribution is actually not spherical. Then the ratio between the relative 
accelerations among LISA test masses due respectively to dust and LDM is 
simply the ratio between the corresponding densities at 1 AU. 
If the LDM value is close to the average GDM value 
$\rho_\mathrm{GDM}= 0.3$ GeV/cm$^3$ = $5 \times 10^{-22}$ kg/m$^3$, 
we have that the effect of interplanetary dust on  breathing motions 
within LISA is of the order of 200 times larger than that induced by a  GDM.  
As for disentangling the two contributions, we see that, even if known with 
moderate accuracy, which, as far as it is concerned, would allow
to improve by a large factor the current best limit on GDM.

The average density of electrons, protons
and helium atoms is somewhat below that of a GDM. Its 
changes during Solar flares develop on so short time scales, in
respect to the inverse of the characteristic frequencies quoted above,
to be ineffective in the data analysis.

The perturbations given by Solar System massive bodies are a more
serious issue. A recent calculation of LISA orbits under the
perturbation of the Earth \cite{14} shows large breathing modes of the arms.
The analysis does not go beyond a first
approximation in the ratio between the tidal effect of the Earth in
respect to that of the Sun. Even preliminary (analytical) calculations appear difficult
and lengthy, and it appears to us that numerical methods are
needed. 

   We believe it would be more fruitful to our purposes to
perform such calculations to see the effect on the orbits of rather
the whole Earth-Moon system. Even if the 
breathing effect due to  the Moon are very small it has a definite
characteristic frequency  that is $f_{EM} = 8.3 \times
10^{-7}$ $\mathrm{Hz}$. This is the double of the frequency of 
of
revolution of the Earth-Moon system around its barycenter and it comes out 
as the first non zero term in the perturbation
after the monopole given by the total
Earth-Moon mass at the barycentre of the system. As we have seen above, 
the LDM effect promotes  signal amplitudes 
in the differential breathings at distinctive higher
harmonics, where there is no signal power
in absence of the LDM perturbation. If this will occur 
in a similar way also in the case of the Earth-Moon perturbation then the 
leading LDM  signal may be present at, say $3f_{EM}$, giving a unique 
signature at  frequencies close to where  LISA is sensitive.

An accurate calculation of the differential breathing of LISA due to the combined effects of the Moon 
motion and of the LDM diffuse density would need a full study of the LISA dynamics, aided by numerical 
calculations, which is beyond the scope of the present paper. We are encouraged by the present initial 
results and such a detailed study is under way (we outline its main features in our Conclusions below).

As discussed above the dust effect may well come out to be indistinguishable in signature 
and about 2 orders of magnitude larger than the effect of a GDM diffuse density. Then, if our 
detailed study will have any success, LISA would give limits to an LDM diffuse density at the 
Earth orbit no better than the limits imposed by the accuracy on independently modelling the 
dust distribution, that is somewhat less then 2 orders of magnitude above the GDM diffuse density. 
This is to be compared with the current limits on an LDM coming from the motion of the planets, which, 
as summarized in the Introduction, are some 5 orders of magnitude larger of the GDM value. 
It is worth noting that future improvements from planetary motion observations, if they concern 
planets within 3.5 AU from the Sun, would suffer from the same limitations from the effect of dust as well.

The notion that dark matter should be present diffusely in the Galaxy
rests on the currently accepted cosmological paradigm
enforced in recent years by a wealth of observational data. The value
of the average GDM is $10^5$ higher than the value implied by
cosmological expansion in the cosmological paradigm, but still it is a direct
consequence of that model, when one wants to account for the rotation
velocities of ordinary matter far from the centre of the Galaxy, on
one side, and on the other side one wants to keep Newtonian
Gravitation, as extended by Einstein with General Relativity, to hold
strictly valid at the Galactic distance scale. However it has been
noticed \cite{15} that there are no independent observational grounds for
such an assumption beyond the Solar System distance scale. A variety
of alternative theories of gravitation have been proposed, which,
while keeping valid Newton-Einstein gravity within the Solar System,
allow a mild violation on Galactic and cosmological scale, with the
benefit to make unnecessary the presence of any dark matter (and even
dark energy). A comparative discussion of the situation says that,
within the present limits given by planets motion, no definitive
conclusion can be reached \cite{1},
but the possibilities of a MOND behaviour 
\cite{bek} within actual LISA orbits should be considered.
A gravitational test like the one we
propose here might give clear cut results in this respect too.

%-----------------------------------------------------------------
\section{Conclusions}
\label{sec:Conclusions}
%-----------------------------------------------------------------

The simple model shows the  potential interest of the
basic idea and may open the way to promising
possibilities. 

We need to investigate to what extend the
perturbations from bodies in the Solar System would stimulate 
 components in the
relative motion between the LISA spacecrafts at the characteristic
frequencies of the perturbation itself, at frequency combinations with
the fundamental frequency of revolution around the Sun and 
at higher frequency overtones which would
be distinctively identifiable in the spectrum of the the differential arms
breathing. To explore
this, and then in case to assess the feasibility of detections/upper
limits with LISA,  we need to embed the LISA test masses in
the most accurate model of dynamics of the Solar System available and
solve numerically for LISA orbits with and without LDM. Such a
procedure will also automatically clarify the issue of the
perturbations given by other massive bodies discussed above. The study
should be completed by a careful evaluation of the effects of the
interplanetary dust and of the Solar wind.
Once understood and
accounted for the effects of massive bodies, interplanetary dust and
Solar activity, a last problem will concern to treat the TDI distance
measurements via light propagation in a fully relativistic way \cite{14}.

\ack 

In the early stage MC benefited from discussions with Michele Bonaldi, 
Livia Conti and Stefano Vitale. We are grateful to Fabrizio De 
Marchi for calling attention to the dust problem and to 
 Oliver Jennrich for discussions. MS is supported by the Swiss 
National Science Foundation and by the Tomalla Foundation.

\end{document}